\PassOptionsToPackage{hyphens}{url}
\PassOptionsToPackage{obeyspaces}{url}
\PassOptionsToPackage{spaces}{url}
\PassOptionsToPackage{dvipsnames}{xcolor}

\documentclass[letterpaper,twocolumn,10pt]{article}
\usepackage{usenix2019_v3}

\usepackage[normalem]{ulem}
\usepackage{soul}

\usepackage{flushend}
\usepackage{balance}
\usepackage{multirow}
\usepackage{graphicx}
\usepackage{makecell}
\usepackage{tikz}
\usepackage{environ}
\usepackage{authblk}
\usepackage{titlesec}
\titlespacing{\section}{0pt}{6pt plus 2pt minus 2pt}{4pt plus 2pt minus 2pt}
\titlespacing{\subsection}{0pt}{6pt plus 2pt minus 2pt}{4pt plus 2pt minus 2pt}
\titlespacing{\subsubsection}{0pt}{3pt plus 2pt minus 2pt}{2pt plus 1pt minus 1pt}
\titlespacing{\paragraph}{0pt}{\parskip}{-\parskip}

\newcommand{\new}[1]{#1}
\newenvironment{newenv}{}{}

\newcommand{\ignore}[1]{}

\usepackage[printwatermark]{xwatermark}
\newwatermark[allpages,color=black!100,angle=0,scale=1,xpos=0,ypos=-126]{\small 2022 31th USENIX Security Symposium \\ Accepted version. \url{https://www.usenix.org/conference/usenixsecurity22/presentation/hlavacek}}

\begin{document}

\pagestyle{empty}

\title{Stalloris:\\RPKI Downgrade Attack} %

\author[*$\S$]{Tomas Hlavacek}
\author[*$\S$]{Philipp Jeitner}
\author[*$\S\ddag$]{Donika Mirdita}
\author[*$\S\dag$]{Haya Shulman}
\author[*$\S\ddag$]{Michael Waidner}
\affil[$\S$]{National Research Center for Applied Cybersecurity ATHENE}
\affil[*]{Fraunhofer Institute for Secure Information Technology SIT}
\affil[$\ddag$]{Technische Universität Darmstadt}
\affil[$\dag$]{Goethe-Universität Frankfurt}

\renewcommand\Authands{ and }

\newcommand*\circled[1]{\tikz[baseline=(char.base)]{
            \node[shape=circle,draw,inner sep=0pt,minimum size=12pt] (char) {\small #1};}}

\maketitle

\begin{abstract}
We demonstrate the first downgrade attacks against RPKI. The key design property in RPKI that allows our attacks is the tradeoff between connectivity and security: when networks cannot retrieve RPKI information from publication points, they make routing decisions in BGP without validating RPKI. We exploit this tradeoff to develop attacks that prevent the retrieval of the RPKI objects from the public repositories, thereby disabling RPKI validation and exposing the RPKI-protected networks to prefix hijack attacks.

We demonstrate experimentally that at least 47\% of the public repositories are vulnerable against a specific version of our attacks, a rate-limiting off-path downgrade attack.
We also show that {\em all} the current RPKI relying party implementations are vulnerable to attacks by a malicious publication point. This translates to 20.4\% of the IPv4 address space.

We provide recommendations for preventing our downgrade attacks. However, resolving the fundamental problem is not straightforward: if the relying parties prefer security over connectivity and insist on RPKI validation when ROAs cannot be retrieved, the victim AS may become disconnected from many more networks than just the one that the adversary wishes to hijack. Our work shows that the publication points are a critical infrastructure for Internet connectivity and security. Our main recommendation is therefore that the publication points should be hosted on robust platforms guaranteeing a high degree of connectivity.
\end{abstract}

\section{Introduction}

The BGP (Border Gateway Protocol) connects the different organisational networks called ASes (Autonomous Systems) by propagating information about how to reach the destinations in remote networks. This central role of BGP makes it one of the most critical components of the Internet infrastructure. Correct functionality of BGP is critical not only for connectivity to services, but also for any security and privacy mechanisms in the Internet, such as  PKI and Tor \cite{DBLP:journals/corr/abs-2004-09063,DBLP:conf/uss/Birge-LeeSERM18,brandt2018domain}. Any configuration error or attack can be devastating to the security and stability of the Internet. Unfortunately, the insecurity of BGP is the cause for frequent Internet outages \cite{fb:out,u:tube,indosat:hijack,turkey:hijack} as well as traffic hijacks \cite{mitm:threat,ballani2007study,vervier2015mind,china:telecom}. 

{\bf RPKI aims to prevent prefix hijacks.} To protect against benign misconfigurations and malicious hijacks the IETF standardised RPKI (Resource Public Key Infrastructure) [RFC6480]. RPKI binds IP address' blocks to the ASes that own them using digital signatures. These signed bindings, called ROAs (Route Origin Authorizations), are stored in public RPKI repositories (called publication points) distributed throughout the Internet [RFC6482]. Each AS can validate that an AS that advertises an IP address block in BGP is authorised to do so. This validation, called ROV (Route Origin Validation), is performed by a relying party, which retrieves the ROA objects from the public repositories. With RPKI validation the BGP routers classify each route learned in BGP into one of three possible route validation states: {\em valid}, {\em invalid} or {\em unknown}. BGP routes with corresponding correct ROAs are classified as valid. BGP routes with incorrect ROAs are classified as invalid. For BGP routes where RPKI objects cannot be retrieved, i.e., there is no valid ROA that covers the IP address block, the validation result is {\em unknown}, in which case RPKI is not used for making routing decisions in BGP. 
This means when an ROA for a prefix cannot be retrieved, the current RPKI deployments attempt to preserve reachability even at cost of security. Our work demonstrates that this design choice makes the current RPKI infrastructure vulnerable.
We show how to downgrade the RPKI protection of ASes, by disabling the RPKI validation of prefixes in BGP announcements that they receive, as a result exposing them to BGP prefix hijack attacks. 

{\bf Downgrading RPKI security.} We present the first RPKI downgrade attacks. Our attacks prevent the relying party implementations from connecting to the RPKI publication points and retrieving the ROAs. The relying parties connect to the publication points periodically every refresh interval, which range between 10 minutes and 1 hour, depending on the relying party implementation. We develop an attack methodology that requires an adversary to cause loss of only a handful of packets between a relying party and a repository in a refresh interval. %
If a relying party cannot connect to the repository, once the cached objects are expired it will not apply RPKI validation. Our measurements show that most objects have validity of 24 hours. We therefore develop an iterative attack, which is launched during the refresh intervals of the target relying party, until the cached objects expire. 
Using Routinator\footnote{According to our measurements, Routinator relying party is supported by more than 66\% of the ASes.}, with its 10 minute refresh interval, as an example, the adversary needs to cause loss of ca. 1.2K packets in 24 hours in the worst case (Table \ref{tab:packet_volumes}) until the objects expire, and the validation is not applied for making routing decisions in BGP.

We then develop a methodology for optimising the attack using a malicious publication point, which causes the relying party to stall, reducing the number of attack iterations to 1.

{\bf Low rate blocking.} We develop a technique to cause off-path packet loss that combines a low rate attack methodology of \cite{conf/sigcomm/KuzmanovicK03} with rate limiting mechanism on the servers, for causing off-path packet loss. %
The idea is to send spoofed packets to the target server using an IP address of a victim. This causes the rate limiting mechanism to kick in, as a result the server starts filtering all the packets that originate from that IP address, even the genuine packets sent by the victim. In our attack, the low rate bursts are synchronised with the intervals at which the relying party sends queries to locate the publication points. The spoofed packets trigger rate limiting at the server for that IP address exactly when the packets from the relying party or its DNS resolver arrive. We find that against some repositories as many as 2-3 packets suffice to trigger the rate limit. Against most servers in the Internet just 1K packets suffice to activate the filtering. In our measurements we found 47\% of the publication points to be vulnerable to rate-limiting downgrade attack. This corresponds to 60\% of the RPKI protected IPv4 address space in the Internet.

Rate limiting is just one example method we use for causing off-path packet loss. Packet loss can be inflicted with other techniques, e.g., by filling the IP defragmentation cache with IP fragments, thus preventing genuine packets from being reassembled \cite{KenMog87,gilad2013fragmentation}. 
Furthermore, since RPKI is meant to provide security against attacks by on-path adversaries, downgrading RPKI is also appealing for strong adversaries that control a router or a network which the packets traverse. 

For attack to be effective it needs to be repeated iteratively, until the objects expire. We show how to optimise this. 

{\bf Stalloris attack.} We develop an attack for stalling the relying parties, which allows us to reduce the refresh intervals of a relying party, correspondingly reducing the iterations needed for a successful RPKI downgrade attack we described above. In fact, we show that a combination of Stalloris with just a single iteration of low rate off-path packet loss attack suffices to remove the RPKI validation. The idea behind our Stalloris attack is to create a deep delegation path so that the relying party opens RRDP connections to multiple publication points controlled by the adversary. The attack is inspired by the Slowloris DoS attack against TCP connections \cite{cambiaso2013slow}, where the attacker opens multiple simultaneous HTTP connections to the target, returning slow responses. In contrast just slowing down the responses, in our Stalloris attack the adversary {\em stalls} the relying party, by creating long delegations chains, which the relying party traverses using multiple RRDP connections to the publication points controlled by the adversary. %

\new{Combining our low-rate attack with Stalloris results in a stealthy and efficient RPKI downgrade attack, which is also extremely difficult to detect.} 

{\bf One attack affects multiple networks.} The extent of our downgrade attack against a relying party with a victim repository is devastating for security and stability of {\em all} the networks that depend on that relying party for RPKI validation, as well as for {\em all} the prefixes that are covered by the ROAs in that repository. First, most ROV supporting networks are large tier 1 providers and all major IXPs (Internet Exchange Points). Smaller networks in the Internet perform RPKI ``by-proxy'': by trusting the ROV of the upstream provider or by trusting the ROV performed by the route-server at the IXP\footnote{Using the numbers from \url{https://rov.rpki.net} there are 127 networks that perform ROV: 34 implement the ROV by themselves, while the rest (93 networks) only depend on IXP route-servers to protect them.}. This means that an attack against the relying party at an IXP has immediate implications for all smaller networks that depend on it.

Second, the repositories host multiple ROAs, for many prefixes. Blocking access to one public RPKI repository has implications for all the prefixes that are covered by the ROAs in that repository, and exposes {\em all} those prefixes to hijack attacks. For instance, consider blocking a public RPKI repository of an RIR, such as RIPE NCC, with a root certificate that is used to delegate to multiple lower certificates in a delegation chain.
We quantify the scope of the networks affected by our attacks.

{\bf Ethical considerations.} Our attacks were carried out ethically against a BGP and delegated RPKI infrastructure that we set up. We validated the attacks: (1) against the publication points in the Internet using our own ``victim'' relying party, and (2) against the relying parties using our own ``victim'' publication point. %
\new{Our evaluations are designed according to the thresholds that the operators as well as the previous work consider to be acceptable, and overall we use less traffic volume than the recent off-path attacks \cite{man2020dns,man2021dns}.}

{\bf Our Contributions:} %

$\bullet$ The relying parties use different timing intervals for retrieving the RPKI objects from the publication points. Not knowing these intervals renders not only the off-path attacks impractical, but makes even the on-path Man-in-the-Middle (MitM) attacks extremely challenging. We develop a method to predict the refresh interval of the relying parties, which is the first step towards making such off-path attacks practical. We analyse the relying parties software and demonstrate that we can accurately predict the timing intervals of the relying party software in RPKI protected networks in the Internet. 

$\bullet$ We systematically analyse the interaction between the different components in the RPKI ecosystem and perform a detailed study of the code and the run time behaviour of popular relying party implementations. This allows us to identify vulnerabilities that expose to our attacks. Our key observations lead to inference of the relying parties' behaviour in the Internet as well as to attacks for stalling their performance. We use these insights to develop an off-path methodology for disabling validation of RPKI supporting ASes. %

$\bullet$ We set up an infrastructure for carrying out measurements of RPKI and conduct an extensive study of the RPKI supporting networks (the relying parties and the repositories). Our measurements provide new insights on the rate limiting in the public repositories and in the nameservers of their corresponding domains, as well as on the retrieval, validation and query behaviour of the relying parties and the DNS resolvers that they use. We find that all the relying party implementations in the Internet have properties that expose them to downgrade attacks. We also find that 47\% of the publication points are vulnerable to the off-path packet blocking attacks via rate-limiting, hence their RPKI protection can be downgraded. This is particularly critical for the 77\% of the publication points that are vulnerable to BGP sub-prefix hijack attacks.

$\bullet$  We demonstrate an example attack against one European IXP, motivating the choice of such attacks against the IXPs.

$\bullet$ We provide recommendations for mitigating our attacks. Nevertheless, the core problem with the ``unknown' RPKI validation status is fundamental and cannot be easily resolved. If the networks insisted on strict validation, classifying the result of missing ROAs as ``invalid'', our attack would block connectivity for those networks to all the prefixes in the unreachable publication point. Hence our main recommendations is that the publication points should be more resilient and should be hosted on stable infrastructure, such as cloud platforms.

{\bf Organisation.} We provide a study of the RPKI ecosystem in Section \ref{sc:dataset}. In Section \ref{sc:analysis} we analyse the behaviour of the relying parties. We introduce our off-path packet loss method and its evaluation against RPKI infrastructure in Section \ref{sc:block}. We present the downgrade attack in Section \ref{sc:downgrade}. We explain how to find the RPKI infrastructure of a target victim AS in Section \ref{sc:find} and describe an attack against IXP in Section \ref{sc:decix}. In Section \ref{sc:mitigations} we explain our recommendations for mitigations. We review related work in Section \ref{sc:works} and conclude in Section \ref{sc:conc}. \new{We provide background on  RPKI in Section \ref{sc:background} in Appendix.}

\section{RPKI Ecosystem}\label{sc:dataset}\label{sc:dataset-pp}\label{sc:dataset-rp}

We performed a study of RPKI deployment in the Internet using a BGP and RPKI infrastructure that we set up. This study forms the datasets of publication points (with their domains and nameservers) and the relying parties (with their DNS resolvers) that we evaluate throughout this work. 

{\bf Measurement infrastructure.} We set up a platform for measuring the RPKI ecosystem. To create our platform we register an LIR account with RIPE NCC and purchase an AS under RIPE NCC and prefixes. We connect our AS to DE-CIX internet exchange point in Frankfurt. We also protect our AS with RPKI. For that purpose we create our own Certificate Authority (CA) and a publication point. %

{\bf Publication points and nameservers dataset.} We extract the list of all active RPKI publication points (aka public repositories) from our running relying party. This list gives us 45 publication point domains \new{resolving to 61 different addresses} and 124 nameservers.

{\bf Relying parties and resolvers dataset.} We study the relying party software in all the RPKI validating networks to understand the refresh interval, as well as the query behaviour. We collect a dataset of relying parties in all RPKI validating networks. Since the query behaviour of a relying party is also related to a DNS resolver it uses to find the publication point, we also collect all the recursive DNS resolvers used by the relying parties.

We find 2699 relying parties (RPs).
We also measure how many different resolvers are used by each RP. We evaluate the number of different ASes in which the resolvers of the relying parties are located. Out of the 2699 RPs, 2163 (80.1\%) were only using resolvers from a single AS, and 1117 (41.4\%) were using only a single resolver. Furthermore, 1984 RPs (73.5\%) used 5 or less resolvers.
We also measured a significant concentration of resolvers: 45.7\% RPs use resolvers operated by Google and 6.7\% RPs use resolvers operated by Cloudflare. %
On the other side, 1273 (47.2\%) RPs were using a resolver in their own AS\footnote{Note that since some RPs use multiple resolvers, it might be that a RP uses both, resolver from its own AS as well as some other ASes.}.

\section{Analysis of Relying Party Behaviour}\label{sc:analysis}

One of the key components of our downgrade attack is the inference of the RPKI objects retrieval during refresh intervals by the relying party software. We first perform a study of retrieval periods by relying party implementations via code analysis and dynamic executions. We explain this behaviour with the most popular Routinator relying party implementation (the relying party developed by NLnetlabs). We then provide an analysis of the retrieval interval of the other (less popular) relying party implementations. %

\subsection{Refresh Interval in Routinator}\label{ssc:routinator}

\begin{figure}[t!]
    \centering
    \includegraphics[width=0.35\textwidth]{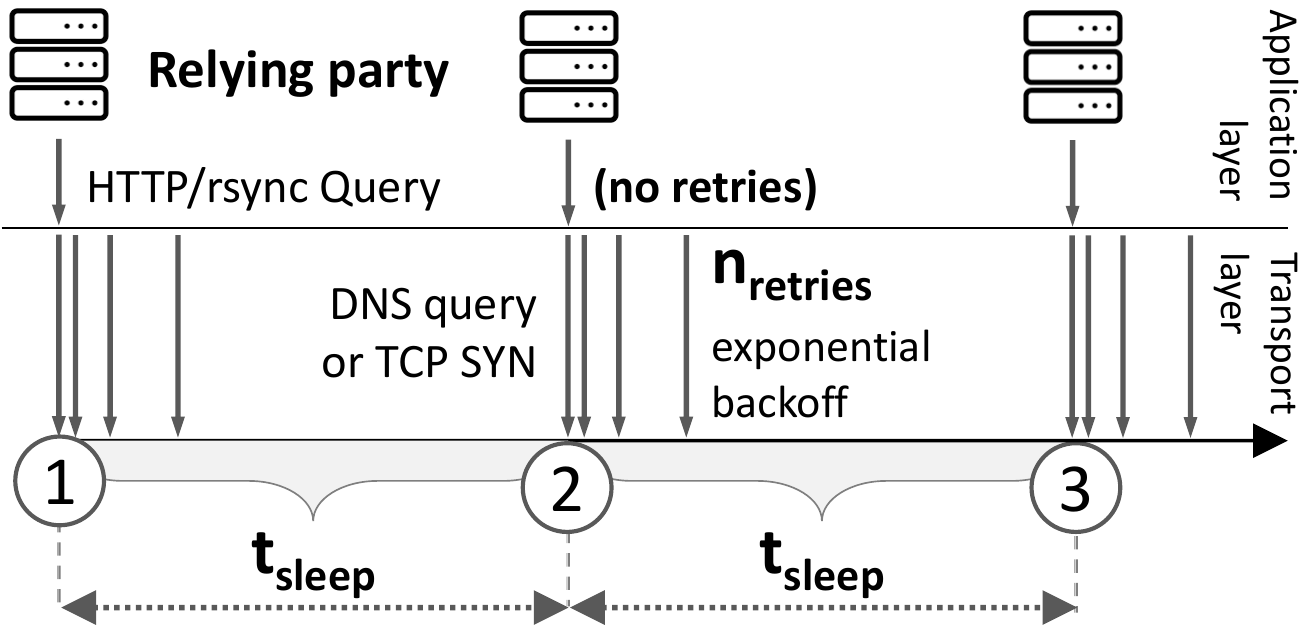}
    \caption{Query behaviour of relying parties.}
    \label{fig:query_behaviour}
    \vspace{-10pt}
\end{figure}

Our main observation is that the validation by the relying party executes sequentially in periods. This conclusion draws on the code analysis of the main functionality responsible for timing of the validation, based on \texttt{PayloadHistory::refresh\_wait()} method in {\it payload.rs} and it is used in Server class in {\it operation.rs}. %
We illustrate the query behaviour of the relying party (with a use case of Routinator) to the publication point in Figure \ref{fig:query_behaviour}. 

There are multiple rendezvous points in the class Server and in the nested code used by this class, that depend on multiple instances of various synchronisation primitives. Moreover, the RTR server is asynchronous IO-bound service that runs in a separate event-loop using Rust Tokio library. The entry point to the asynchronous RTR code is in {\it rtr.rs} file. %

(1) As we see, the refresh timing can be affected by multiple locks, including the one that depends on asynchronous IO-bound RTR server that share data structures with the validation code. We tested ordinary use cases to quantify likelihood of triggering of locking delays and we conclude that the internal delays caused by locking are marginal. This observation improves the accuracy of our method of timing prediction despite the internal complexity of the Routinator software. %

(2)  Under normal circumstances (no network outages nor publication point timeouts) the refresh takes 15-45 seconds, the first 5 seconds are spent by scanning the local cache, which depends on the local storage speed. Because the validation runs sequentially (see (1)) the length of the RRDP and rsync refresh depends on the response times of the servers.

When a publication point is unreachable and the TCP/IP stack fails to signal the unreachability over ICMP, there are multiple timeouts, the most important ones for us are the RRDP and rsync timeouts, both 300 seconds. This provides an upper bound on the delay that can be caused by a single unresponsive publication point. %

\begin{figure}[h!]
    \centering
    \includegraphics[width=0.45\textwidth]{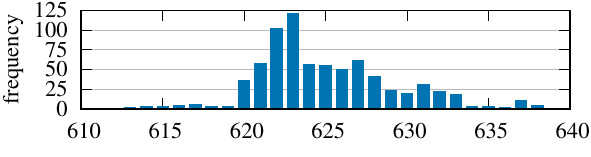}
    \vspace{-10pt}
    \caption{Routinator timing in seconds {\footnotesize  ($\mu=625$s, $\sigma=4.3$s)}.}
    \label{fig:routinator_timing}
    \vspace{-10pt}
\end{figure}

(3) The time of refresh does not count in the refresh-period. That is the reason why the mean time between refreshes is 625 seconds. 
We also measured distribution of the delays between requests to RRDP and rsync publication points, and plot the results with the histogram in Figure \ref{fig:routinator_timing}.

\subsection{Refresh Interval in Other Relying Parties}

In addition to our in-depth analysis of Routinator, we analyse the query behaviour of all major relying party (RP) implementations.

\begin{figure}[t!]
    \centering
    \includegraphics[width=0.45\textwidth]{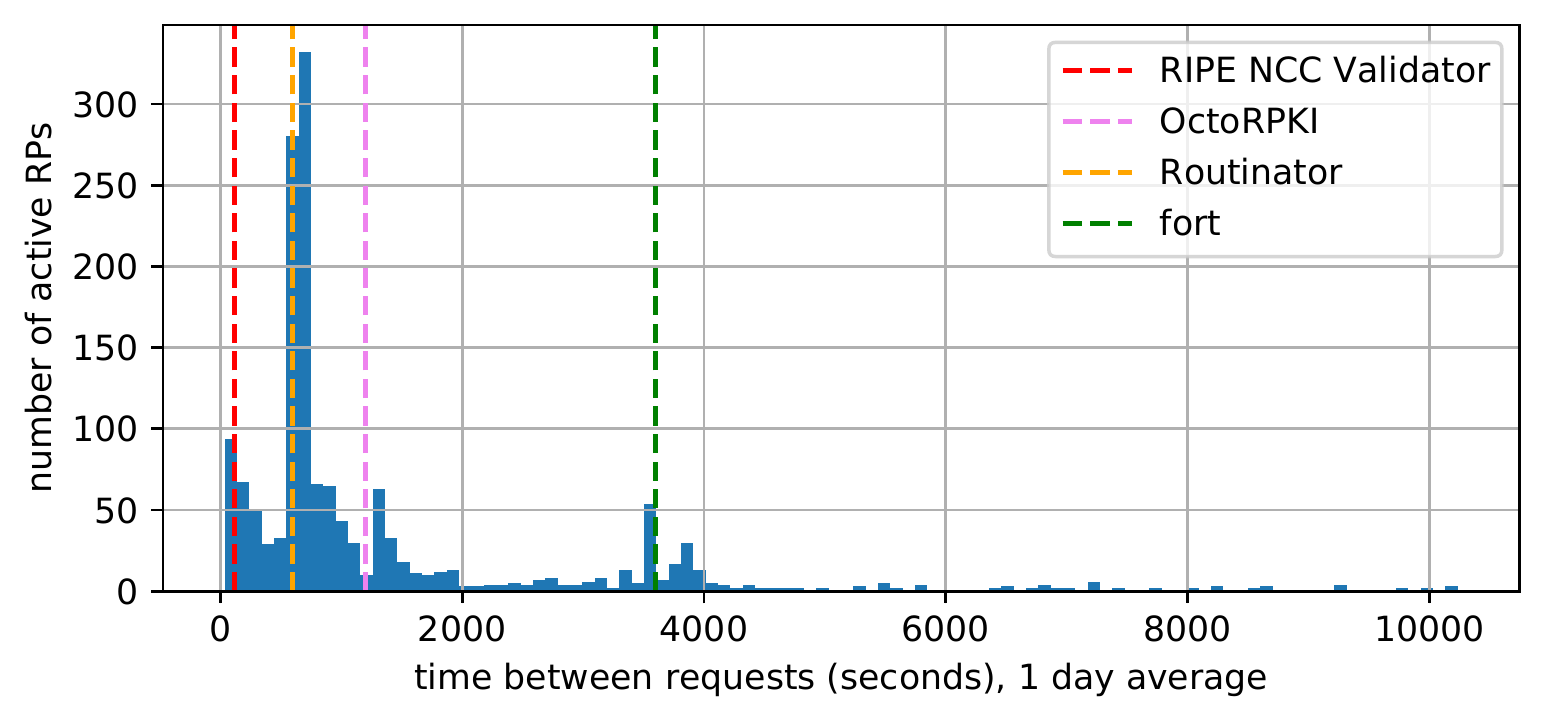}
    \vspace{-10pt}
    \caption{Distribution of time between requests at our PP.} %
    \label{fig:rp-schedules}
\end{figure}

To measure the frequency of the request schedules of the relying parties, we use our relying parties dataset, described in Section~\ref{sc:dataset-rp}. We log all the requests at our publication point and calculate statistics on the time-span between consecutive queries sent by each relying party (differentiated by IP address). We then
correlate the results with a source code analysis of the relying parties and use the User-agent header sent in the case of RRDP as a source of information about the relying party implementation. We show the distribution of the time between 2 consecutive requests by the same relying party in Figure~\ref{fig:rp-schedules}.

In Table~\ref{tab:rp-schedmed} we show the mean time and standard deviation between 2 consecutive queries for each major relying party implementation along with a theoretical value obtained from the source code analysis. Our results show that the mean experimental values are in the same range as the source code values, but there is a significant deviation. As we showed for Routinator, this deviation comes from the fact that Routinator does not account for the time the RPKI update takes in the sleeping interval and as such we conclude that this deviation is the deviation of the duration of the update for different relying parties. As shown in Figure~\ref{fig:routinator_timing}, the deviation is much smaller when only measuring a single relying party.

\begin{table}[h!]
    \renewcommand{\arraystretch}{0.8}
    \setlength{\tabcolsep}{3pt}
    \footnotesize
    \centering
    \begin{tabular}{c|c|c|c}
Relying party software & \makecell{default refresh\\interval (from code)} & \makecell{measured \\ $\mu$ (s)} & \makecell{$\sigma$} \\
\hline
 Routinator & 600 & 686.5 & 132.9 \\
 fort & 3600 & 3634.5 & 86.1 \\
 Cloudflare-RRDP-OctoRPKI & 1200 & 1465.0 & 167.8 \\
 RIPE NCC RPKI Validator & 120 & 158.0 & 133.0 %
    \end{tabular}
    \caption{Measured time between requests $t_{sleep}$.}
    \label{tab:rp-schedmed}
    \vspace{-10pt}
\end{table}

\section{Off-Path Packet Blocking}\label{sc:block}
An important aspect of our attack is to block packets remotely. 
We exploit rate limiting of servers to block responses to the relying party, ``simulating'' packet loss. The target servers are publication points as well as the DNS nameservers which the DNS resolvers of the relying parties query to find the publication points.

\begin{table}[t!]
\renewcommand{\arraystretch}{0.9}
{\footnotesize
    \centering

\setlength{\tabcolsep}{2.6pt}
\begin{tabular}{c|c|c|c|c|c|c|c|c|c}
responses/s & 3 & 4 & 445 & 1137 & 1142 & 1146 & 1146 & 1207 & 1212 \\ 
\hline
answers/s & 3 & 2 & 82 & 1137 & 1142 & 1146 & 1146 & 1207 & 1212 \\
\hline
\hline
responses/s & 1223 & 1287 & 1288 & 1296 & 1308 & 1309 & 1520 & 1642   & 2621\\
\hline
answers/s & 1223 & 1287 & 1288 & 1296 & 1308 & 1309 & 12   & 1301   & 1236\\
\hline
\hline
responses/s & 3248 & 3248 & 4000 & $\infty$ & $\infty$ & $\infty$ & $\infty$ & - & - \\
\hline
answers/s & 16 & 16 & 88 & 3 & 7 & 7 & 7 & - & - \\
\end{tabular}

    \caption{DNS response-rate-limiting (RRL) in PP domains. }
    \label{tab:pp_ns_rate_limits}}
\end{table}

\subsection{Rate-limit in DNS}
\label{sec:measure-ns-ratelimiting}\label{sec:measure-resolver-ratelimiting}

DNS Response Rate Limiting (DNS RRL) \cite{dnsrrl} is a technique employed by recursive resolvers and authoritative nameservers to reduce the impact of DNS amplification and reflection~\cite{rozekrans2013defending} attacks. When a DNS server uses DNS RRL, it monitors the rate of requests coming from an individual client (or address prefix), typically with a token-bucket scheme \cite{dnsrrl-bind}, and will stop answering queries originating from this client when the volume reaches a configured rate-limit.

We measure the rate limit in nameservers in the domains of publication points, and in resolvers. Since public DNS resolvers are often used by the resolvers of relying parties, triggering rate limit in public DNS resolvers has a similar effect on the losses to the resolver, like the rate limit in nameservers.

\subsubsection{DNS Nameservers}
When a DNS server finds a client has reached its preconfigured rate-limit, it has several methods of denying service to the client. In practice, we find 2 common types of rate-limiting in DNS servers, which are often combined and used in an escalating way: (1)~Dropping some (or no) responses\footnote{In Bind9, this first limit only applies to "similar" queries, meaning that queries which ask for a different domain or record type are not affected. We do not consider this because in our case the attacker knows the domain and can therefore spoof requests asking for the correct domain and type.}, but respond to the rest with a empty DNS response with a TC bit \cite{RFC1035} set, which indicates to the client that it should reconnect over TCP. (2)~Dropping all the responses to a client and do not answer any requests until there are more response-tokens available.

The limits for both approaches can often be separately configured in DNS servers\footnote{In Bind9 with the 'responses-per-second' and 'all-per-second' settings.} and often, the first (1) method is used with a low rate-limit, while the second (2) is used with a higher limit \cite{bind-docs}.

{\bf Trigger rate-limit to block packets.} To exploit the rate limiting feature to deny service to a third party client, an attacker can abuse the feature by sending spoofed requests with the address of the client to the server. Upon reaching the limit, the server will either send only truncated responses or no responses at all, which affects all packets with the address of the client. When sending back truncated responses, the client has the option to still get served by connecting back via TCP, so the attack has no further consequences\footnote{We tested the support of TCP in the resolvers of relying parties and found that almost all resolvers where able to connect back using TCP, rendering such an attack pointless.}. However, when the server uses the second technique and stops answering, the client will instead assume that the server is unresponsive and fail the resolution, which will cause the client, that requested the resolution, to be unable to connect to the service at the requested domain.

\begin{table}[t!]
    \renewcommand{\arraystretch}{0.8}
{\footnotesize
    \centering
    \begin{tabular}{r|c|c|c}
Provider & Address & \makecell{responses\\per sec} & \makecell{answers\\per sec} \\
\hline
Google & 8.8.8.8 & 1500 & 500 \\
Cloudflare & 1.1.1.1 & 1000 & - %
    \end{tabular}
    \caption{Rate-limiting in public resolvers.}
    \label{tab:public_resolver_limits}}
    \vspace{-10pt}
\end{table}

{\bf Measurement of rate limit in nameservers.} We test the rate-limiting in all nameservers responsible for publication points in our dataset, by probing the nameservers with different query rates from 1/sec to 10000/sec for short periods of time. We monitor the rate of responses and check whether the responses are "slipped", i.e., do not contain answers but only have the TC bit set. For each namesever, we then calculate the effective number of responses and answers (i.e., responses which contain answers and do not have the TC bit set) per second by dividing the total number of responses by the test duration. We consider a publication point vulnerable if all the nameservers in its domain enforce a rate-limit. We list the maximum number across all the nameservers per publication point in Table~\ref{tab:pp_ns_rate_limits} for all domains where such a limit could be determined\footnote{\new{We removed the domain names for anonymity reasons.}}.

In total, we find 25 RPKI repositories where all the nameservers enforce rate-limiting. Out of these, we consider 21 vulnerable, since they will stop sending responses at some point. The other 4 only "slip" responses, meaning that an attack is not practical assuming the resolver is able to connect via TCP. Furthermore 2 out of the 21,  are particularly vulnerable because of the very low rate-limit of only 3-4 responses per second, which is very easy for an adversary to trigger.

\subsubsection{Public DNS resolvers}

We also test rate-limiting applied on client queries in the 2 major public resolvers which were used by the relying parties in our dataset: Google (8.8.8.8) and Cloudflare (1.1.1.1); 45.7\% of the relying parties use Google and 6.7\% Cloudflare.

We find that operators of both resolvers enforce rate-limits, which we list in Table~\ref{tab:public_resolver_limits}: Google uses a rate-limit of 500 queries/s at which point some of the responses are slipped (no answer, TC bit set) and a total response limit of 1500/s at which point no answer is sent at all. Cloudflare uses a total response limit around 1000/s but did not use the slip mechanism. This numbers are inline with the documentation provided by Google\footnote{Google specifies a limit of 1500 queries/sec for their public resolver: \url{https://developers.google.com/speed/public-dns/docs/isp}}.

Hence, adversaries can block responses to a relying party by exploiting the rate-limit in public resolvers. Since both resolver infrastructures are anycast-routed, the adversary needs to select an attack node at the same anycast-instance of the resolver platform. %

\begin{figure}
    \centering
    \includegraphics[width=0.47\textwidth]{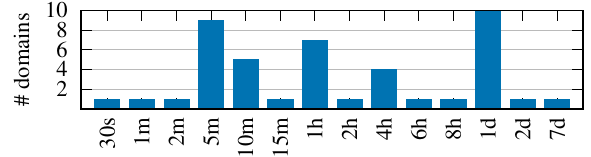}
    \vspace{-5pt}
    \caption{A Record TTL distribution for Publication Points.}
    \label{fig:a_ttl}
    \vspace{-10pt}
\end{figure}

\subsubsection{Caching and TTL}

When exploiting the rate-limiting in nameservers to make domain resolution unavailable, the adversary has to wait for any cached records to be removed from the resolver's cache. We show experimentally that the TTL of publication point URLs are low: over 93\% of publication points have TTLs of 5 minutes and only 22\% of resource records are 1 day or longer. The distribution of the TTL values for A records for publication points is plotted in Figure \ref{fig:a_ttl}. %

\begin{figure}[t!]
    \centering
    \includegraphics[width=0.47\textwidth]{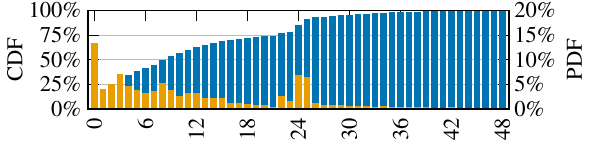} \\
    \vspace{-8pt}
    {\footnotesize \new{Yellow: PDF, Blue: CDF.}}
    \vspace{-5pt}
    \caption{Request intervals for PP resolvers in hours. }
    \label{fig:req_dist}
    \vspace{-10pt}
\end{figure}

Furthermore, as we show, even the instances with high TTLs are limited by the cap of the resolvers. In our experiment with the resolvers of the relying parties we set the TTL of our record to 48h, then trigger queries to that record and monitor traffic at our nameservers. In Figure \ref{fig:req_dist} we show the request frequency for our resource record in a 48h observation period \new{where the yellow bars show the frequency and the blue bars show the cumulative frequency of all the values below a given point.} According to the graph most resolvers have their own individual maximum cache time for records. The median value of the measured caches is 8h. This observation indicates that even if publication points have large TTLs, the maximum cache time parameter in resolvers would ignore high TTLs and instead use smaller values. %

\subsection{Rate-limit in Publication Points}
\label{sec:measure-pp-ratelimiting}
Additionally to rate-limiting in DNS servers, many operators also use rate-limiting in other services, either to protect the service against application-level DoS \cite{molsa2004effectiveness}, TCP-Syn-Flooding attacks \cite{ciscopress}, or to slow down port-scanning attempts against the service \cite{nmapscanning}. Such rate-limits are typically implemented in Firewalls \cite{iptables-hashlimit} which monitor the connection attempts by individual clients and drop any requests above a certain preconfigured limit.

As this filtering happens right at the connection stage of a TCP connection, attackers can exploit this to consume all the connection attempts allowed for a client during a certain time-period by sending a stream of spoofed TCP SYN packets to the service. Similarly to the second method of rate-limiting in DNS, this will lead the client to assume that the service is unresponsive and eventually fail the connection attempt. 

{\bf Measuring rate limit in PP.} We test rate-limiting in publication points in terms of the number of TCP SYN packets answered. We probe each publication point IP address in our dataset with different rates of TCP SYN packets to the publication point TCP port and measure the amount of answers received. To reduce the load on the servers caused by this measurement, we limit our measurement to 6 seconds. If the rate of answers sent back to our test machine is significantly below the number of requests (TCP SYNs) sent, we consider this answer rate as the rate limit of the server. We consider a rate-limit to be hit when we see a response rate of under 90\% of the rate of TCP SYN packets sent during our 6 second test interval.

A rate limit was hit in 8 servers out of 61 tested, listed in Table~\ref{tab:tcp_syn_rate_limits}. Out of these 8 servers, 7 where solely responsible for serving the publication point URL and thus are vulnerable to the low-rate attack.
Similar to our evaluation of rate-limiting in DNS nameservers, we observed 2 publications points where the rate limit is particularly low. %

\begin{table}[h!]
 \renewcommand{\arraystretch}{0.8}
{\footnotesize
    \centering

    \setlength{\tabcolsep}{4pt}
\begin{tabular}{c||c|c||c|c|c|c|c|c}

rate-limit/s & 10    & 60    & 2326  & 4667  & 4805  & 5394  & 5838  & 8842  \\
\hline
only server & yes   & yes    & yes  & yes  & yes  & yes  & yes  & no  \\
\end{tabular}

    \caption{TCP SYN rate-limits in publication points.  }
    \label{tab:tcp_syn_rate_limits}}
    \vspace{-10pt}
\end{table}

\begin{newenv}
\subsection{Ethical considerations for RRL scans}
Our experiments for determining the DNS RRL and TCP SYN rate-limits are comparable to other experiments measuring rate-limiting in remote servers \cite{man2020dns}. In addition, we communicated with DNS operators to coordinate acceptable rates that do not affect the operation. This includes CZ.NIC ccTLD for cz. that manages more than 1.4 M domains, a major IXP NIX.CZ that operate J- and K-root anycast DNS servers, WebGlobe (AS29134) - a large ISP that operates authoritative DNS servers for more than 100k domains, and TI AG (AS29655) - a major ISP that operate recursive DNS servers for large FTTH metro network. From our survey of network operators we found that "1000 DNS requests/sec" is not causing impact on authoritative DNS servers in the Internet, so running an experiment that does not exceed this volume of requests should not cause operational problems. We tested the low rate attack using low rates and our tests lasted only 6 seconds in total. Given that modern nameservers are able to serve 500000 answers per second without any problems \cite{knot-benchmark}, at peak we have not even used 2\% of the capacity of such a nameserver for 6 seconds. We limit the experiment to 6 seconds per server, so that it is shorter than the connection timeout of a typical client. %
In our queries we use the same query (domain+type) and do not ask for DNSSEC records in order not to increase the load. %
Finally, we only conduct the experiment against one server at a time, so that the fallback servers are not affected.

\end{newenv}

\section{RPKI Downgrade Attacks}\label{sc:downgrade}

During the ROV a relying party uses information in the RPKI to make routing decisions in BGP, i.e., to decide if to accept an IP prefix and an origin AS pair. A relying party queries the domain for each repository to locate the server that hosts the repository, then connects to it to retrieve the RPKI objects over RRDP or rsync. The valid ROAs are then used to classify the AS and prefix pairs in each BGP announcement as valid, invalid or unknown. If the validation results in status unknown, RPKI validation is not applied. In this section we explain our RPKI downgrade attack, in order to hijack the BGP prefix of the target AS. Our attack consists of low-rate packet loss attack against the publication points and of Stalloris attack against the relying party. %
We then evaluate and analyse the success probability of the attack against RPKI-protected networks in the Internet.

\subsection{Low-Rate Attack}\label{sc:attack-overview}

{\bf Find relying party of victim AS.} Assume that the two victims are AS $T$ and AS $B$. The adversary wishes to cause AS $T$ accept a hijacking BGP announcement for AS $B$. Both AS $T$ and AS $B$ deployed RPKI. The adversary finds the relying party of $T$ and the DNS resolver that the relying party on the victim AS $T$ is using. The adversary also finds the public repository (aka publication point) of $B$, which serves the RPKI information for the prefix of $B$. We show how to identify the relying party in Section~\ref{sc:find} and the resolver in Section~\ref{sc:dataset-rp}. 

{\bf Selectively dropping packets.} Once the victim relying party of network $T$ and the public repository of network $B$ are identified, the attack consists of preventing the relying party from communicating with the target public repository. This can be achieved by dropping the packets exchanged between the relying party and the public repository. For instance, the adversary can continually flood the communication path between the relying party and the repository. Such an attack requires substantial adversary capabilities and can be exposed since such traffic volumes would also disrupt other services. 
\begin{figure}[t!]
    \centering
    \includegraphics[width=0.4\textwidth]{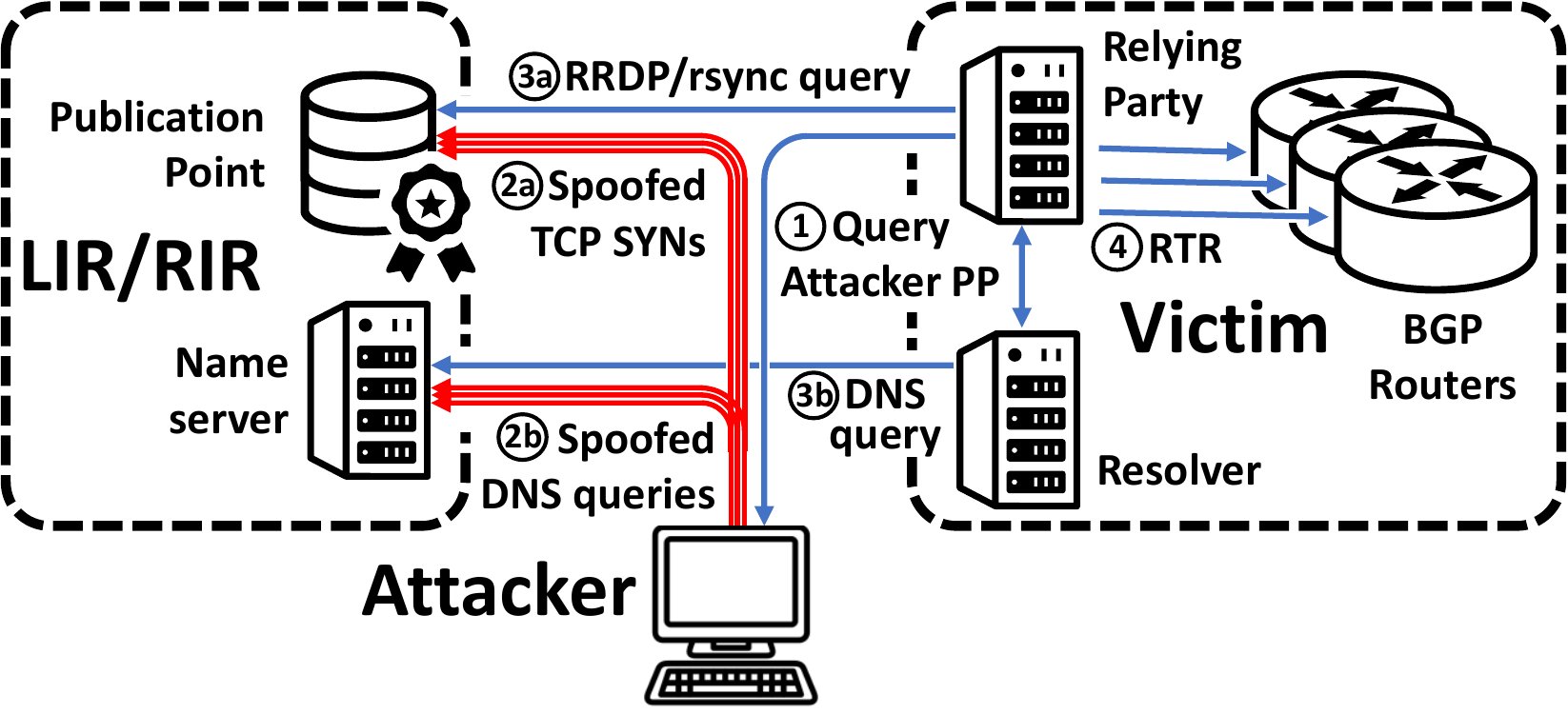}
    \caption{Attack overview.}
    \label{fig:overeview}
    \vspace{-10pt}
    \end{figure}
    
We use a rate-limiting based off-path attack methodology we showed in Sections~\ref{sec:measure-ns-ratelimiting} and \ref{sec:measure-pp-ratelimiting} in order to drop packets exchanged between the relying party and the repository. We experimentally found that many of the publication points and the nameservers support rate limiting, see Tables \ref{tab:pp_ns_rate_limits} and \ref{tab:tcp_syn_rate_limits}. In addition, above 50\% of the resolvers of the relying parties use public upstream resolvers, which also enforce rate limiting, hence extending the attack surface, Table \ref{tab:public_resolver_limits}. Our measurements in Section \ref{sc:block} demonstrate that the rates for blocking communication are moderate hence allow practical attacks. However, since the attacker does not know when the packets are exchanged, it would require to constantly run the low rate attack exploiting rate limiting, which makes the overall attack not practical. We develop an adversarial strategy to synchronise the low rate attack to the ``refresh intervals'' of the relying party, during which the relying party queries the public repository. 

{\bf Predict the query interval of the target.} In order to launch network attacks to disrupt connectivity the adversary needs to predict when the relying party initiates the query interval. At that point the relying party starts connecting to the repositories to request the RPKI objects. This phase is accompanied with DNS queries, to locate the repository, and connection establishment with RRDP or rsync. Predicting this phase is believed to be difficult, especially for off-path adversaries, since they are not located on the path, hence cannot see any communication exchanged between the relying party and the public repository. %

To be able to predict the retrieval interval our adversary sets up the BGP and RPKI infrastructure, in delegated RPKI setup (see description in Section \ref{sc:dataset}). This enables our adversary to directly receive communication from any relying party in the Internet. In particular, also from the victim AS (step \circled{1} in Figure \ref{fig:overeview}). Using our analysis in Section \ref{sc:analysis} we show that this allows the adversary to predict the refresh interval since the queries to {\em all} the repositories are sent concurrently in one bulk. Hence, once queries from the relying party of AS $T$ arrive at the nameserver or the repository of the adversary, the adversary can calculate the next query interval and synchronise its attack accordingly. This ability to anticipate the queries allows us to significantly reduce the volume of the attack while improving its effectiveness. In Section \ref{sc:success:analysis} we show that this approach allows to launch practical attacks against targets in the Internet. Without the ability to synchronise the attack the only strategy of the adversary would have been to continually prevent communication between the relying party of $T$ and the target repository of $B$. 

\begin{figure}[t!]
    \centering
    \includegraphics[width=0.45\textwidth]{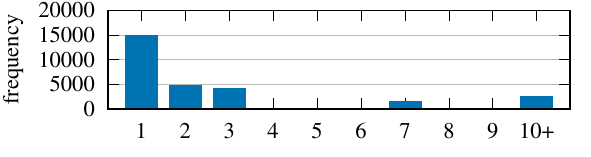}
    \vspace{-10pt}
    \caption{RPKI manifest validity period in days.} %
    \label{fig:mft_validity}
    \vspace{-10pt}
\end{figure}

{\bf Attack until RPKI status ``unknown''.} As soon as the query interval is found, the adversary needs to launch the attack once in every refresh interval until the RPKI validation is not be applied for making routing decisions (step \circled{2a}/\circled{2b} in Figure \ref{fig:overeview}). When the relying party cannot reach the publication server (step \circled{3a}/\circled{3b}, Figure \ref{fig:overeview}) and does not have the RPKI objects for that AS in its cache, it will not be able to perform the ROV validation. The RP will eventually change the RPKI state of the target prefix to unknown and will update the BGP routes (step \circled{4} in Figure \ref{fig:overeview}). If the objects are cached, the adversary needs to wait until they expire. 

\begin{figure}[t!]
    \centering
    \includegraphics[width=0.45\textwidth]{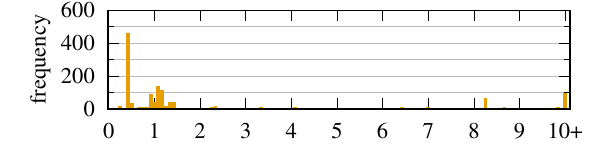}
    \vspace{-10pt}
    \caption{ROA objects validity period in years.} 
    \label{fig:validity:roa}
    \vspace{-10pt}
\end{figure}

{\bf How long does it take for cached objects to expire?} Our measurements show that ROA objects are typically valid for long time periods. Our measurements show that the median validity of ROA objects is 545 days; see validity distribution of ROA objects in Figure \ref{fig:validity:roa}. However, for efficiency purposes (to reduce the sizes of the revocation lists and the volume of the BGP updates), the standard requires the relying parties to {\em only} check the manifests. The publication servers and the relying parties use the manifest files to signal changes in the repositories and to avoid certain types of object deletion and substitution attacks with older (but still valid) objects. All objects of stale manifest files and all the subtrees of these objects should be treated as ``suspicious'' as instructed by [RFC6486]. We checked the behaviour of all the RPKI validator implementations (the list is in Figure \ref{tab:rp-schedmed}, Section \ref{sc:analysis}) and found that they are all [RFC6486] standard compliant and strict towards ``suspicious'' objects in their default settings, i.e., they discard all the ``suspicious'' objects and the entire subtrees, for which there are stale manifest files, and exclude them from the RPKI validation.

\textbf{How long does it take for manifests to expire?} To answer this question we performed a measurement of the validity period of the manifest files [RFC6486]. 
Each publication server generally needs at least one manifest, but the structure of the hosted RPKI system operated by RIRs provide a separate space and manifest file for each hosted customer. We downloaded data of all the active RPKI publication servers on July 15, 2021 and obtained 28330 valid manifest files. We compute the distribution of the validity period lengths and plot the results in Figure \ref{fig:mft_validity}. The results are that there are 15018 (53.01\%) manifest files with validity $\leq$24 hours, 19836 (70.02\%) that are valid for $\leq$48 hours, 24038 (84.85\%) valid $\leq$96 hours and 25690 (90.68\%) manifest files are valid for $\leq$15 days. Hence, in the worst case the adversary needs to wait just one day until it can complete the attack on 53.01\% of the publication servers. However, in practice, \new{manifest files are not generated on the fly when they are downloaded by an RP, but are re-generated periodically by publication points. In Krill, an open source RPKI CA with a built in publication point, we find that manifest files are re-generated once their remaining validity period is below 6 hours. This means an attacker can time his attack to a point in time when a manifest file is about to reach 6 hours of remaining validity so he only need to wait the remaining 6 hours for it to expire.}

Once the manifest expires the RPKI status for that AS will result in status {\tt unknown}, instead of {\tt invalid}. Consequently, when making routing decisions and updating the forwarding tables the border routers on that AS will not apply RPKI for validating the BGP announcements of the origin AS whose prefix the adversary wishes to hijack. The adversary can issue a bogus BGP announcement to hijack the prefix of victim $B$. Depending on the local preferences of $T$, if it accepts the hijacking BGP announcement for $B$, all the communication of the hijacked origin $B$ will be sent through the adversary. This is despite the fact that both $T$ and $B$ are RPKI-protected.

How many attack iterations are required to cause status unknown? As an example we take manifest validity of one day (24 hours) and we assume that the AS $T$ uses the Routinator relying party. Routinator uses 10 minutes refresh intervals, which results in 144 attack iterations until the relying party of $T$ does not apply RPKI validation, exposing itself to accepting hijacked BGP announcements. Although in most cases this allows practical attacks (see analysis in Section \ref{sc:success:analysis}) we can further significantly reduce the attack complexity with an optimisation that we develop. The optimisation allows to reduce the attack iterations to as little as 1.

\subsection{Stalloris Attack}\label{sc:optimization-delegation}

We develop an attack to stall the relying party. This attack can be used in combination with the low rate attack resulting in overall optimisation of the iterations required to downgrade RPKI. 

\begin{figure}[t!]
    \centering
    \includegraphics[width=0.35\textwidth]{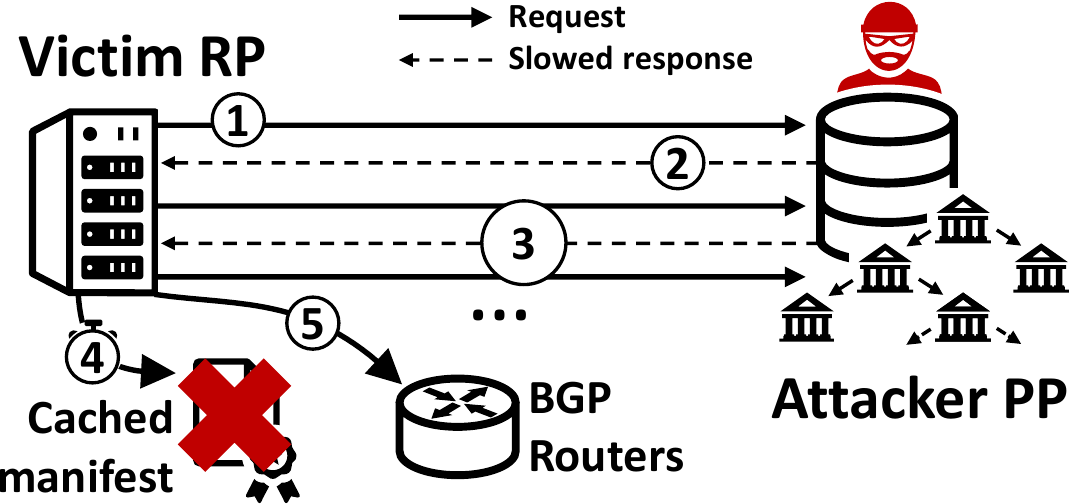}
    \vspace{-5pt}
    \caption{\new{Stalloris attack.}}
    \label{fig:stalloris}
\end{figure}

{\bf Unlimited delegation paths.} In our code analysis of the relying party implementations we noticed that the relying parties do not check the delegation path (length, depth or even both). We use this observation to encode arbitrary long delegation paths. First, quick background information. In RPKI it is possible to delegate authority over an IP prefix and an AS number: our adversary delegates the resources to itself, e.g., to another repository under the subdomain that it set up. The relying party starts the RPKI validation from the root, this is the RIR, e.g., RIPE NCC. From there, it follows the delegation to at least one resource pointing at the CA, which points to the publication point. The relying party contacts each publication point and every domain in the delegation chain and downloads the data. When it identifies a new delegation chain, it contacts the publication point, retrieves the data and stores it in the cache -- this occurs within one refresh period. If there is a delegation to the next level, during the next refresh interval, the relying party will contact that publication point and download the certificates over RRDP again. The publication points in a delegation can all be mapped to the same IP address: As long as they are mapped to different subdomains, from the perspective of the relying party these are different publication points, and the relying party will attempt to traverse all of them.

{\bf Stalling relying party.} The idea of the attack is to set up multiple publication points and to construct a long delegation chain, which the adversary provides in an answer to the target relying party, when it queries the publication point controlled by the adversary. The adversary returns slow replies in the RRDP transactions from all the publication points that it controls, in order to degrade the performance of the victim relying party. By slowing down the relying party, the adversary reduces the number of refresh intervals, hence reducing the number of iterations needed to launch the attack. We call this attack Stalloris\footnote{Loris is a slow animal \path{https://en.wikipedia.org/wiki/Loris}.}, since it stalls the relying party, and is inspired by the Slowloris DoS attack \cite{cambiaso2013slow}, where an adversary initiates multiple slow HTTP connections to the victim server.

\begin{newenv}
To increase stealthiness the adversary may limit the attack to some selected victims. To the other relying parties the adversary sends simple and valid answers. The adversary responds with different certificates (with or without downstream delegation) to expose more levels in the delegation chain only to the victim RP and not the rest of the world. Hence only victim RP is affected, but not the others. Detecting our attacks from logs or monitoring is tricky: servers may become unreachable for legitimate reasons, operators cannot differentiate such attacks from legitimate failures. We next explain the steps of the attack.

This attack is illustrated in Figure~\ref{fig:stalloris}: In step (1), the victim RP connects to a adversarial PP to fetch the ROAs. In step (2) the adversarial PP is configured to answer all requests as slow as possible while avoiding errors in the victim RP by limiting the transmission rate and artificially inflating the response size. However, instead of providing the respective ROAs, the adversary sets up multiple layers of sub-delegations to other RPKI CAs which point back to other PPs in the adversary's infrastructure. To chase down the delegation chain, the victim RP therefore has to connect back to another adversarial PP in step (3), which again, responds slowly. This process is repeated by the adversary. In step (4), after some time, the cached manifest file from the victim RP times out because it has not been refreshed due to the RP being stalled in the process of downloading ROAs from the infrastructure of the adversary. Therefore, when the RP performs the next RPKI validation run, it removes the ROAs referenced by the stale manifest and sets the RPKI state of the corresponding prefixes to unknown, communicating these changes to the connected BGP routers in step (5).
\end{newenv}

The Stalloris attack can be used as a supplementary optimisation to our low rate attack, but cannot completely replace it. Since the re-validation occurs after the refresh finishes, for a successful downgrade attack the relying party needs to update the cache once (i.e., successfully retrieve the RPKI objects), then to fail once (due to our low-rate packet blocking method) and only then do we proceed to stalling the relying party (with Stalloris). %
Combination of low-rate blocking attack with Stalloris results in an highly efficient attack; we provide analysis of traffic volume for different attack combinations and configurations of relying party and publication point in Section \ref{sc:success:analysis}.

{\bf Duration of each delegation level.} The processing of each level in a delegation chain is limited by a timeout in the relying party; in Routinator this is 300 seconds, see analysis in Section \ref{ssc:routinator}. Hence, the adversary needs to keep the relying party at each level for at most 300 seconds, otherwise the relying party will timeout. %

{\bf How many levels are needed?} If the adversary creates a delegation chain of 288 levels, it will cause the relying party to be stalled for 300$\cdot$288=86000= 24 hours. The adversary can therefore combine this optimisation with our low rate attack, hence reducing the number of iterations for attack. For instance, if it creates 144 delegation path, it stalls the relying party for 12 hours, and the adversary needs to launch the low rate attack once. Given a 72 level delegation path, the adversary needs to run the low rate attack every 6 hours. 

{\bf Requirements on the adversary.} The adversary needs to control one AS and one /24 prefix, and can delegate them over and over again. It does not actually need to split the prefix in its delegations, and can make a long chain using the same prefix. The adversary also needs to control a CA and a publication point. The adversary needs to control a domain for its publication point, and needs to create multiple subdomains. Each subdomain is used to create a delegation level. All these requirements are straightforward for adversary - we validated this by acquiring and setting up routing and RPKI infrastructure under RIPE NCC. Finally, the adversary needs to generate multiple CA and publication point domains, all can point to the same IP address. This step can be automated with a script.

\begin{newenv}
{\bf Evaluation in RP implementations.}
We evaluate our Stalloris attack against the major RP implementations, by measuring the maximum time the relying parties can be stalled on a single connection by not sending any data (and remaining idle) and throttling the connection in combination with artificially inflating the download size\footnote{We used a 100MB inflated file at 100kbps.}. Additionally, we evaluate the maximum certificate chain depth. The results of our evaluations are listed in Table~\ref{tab:stallingtest}. Our findings are that Routinator, Fort and OctoRPKI can be stalled between 5 to 10 hours per update, by maximising the connection time for each publication point with the maximum chain depth. Note that an adversary can also increase the chain width by delegating to multiple sub-publication points at each level.

\begin{table}[t!]
 \renewcommand{\arraystretch}{0.8}
\begin{newenv}
    \setlength{\tabcolsep}{4.5pt}
\centering
\footnotesize
\begin{tabular}{ c|c|c|c|c|c }
               & \multicolumn{2}{c|}{HTTPS} & Rsync       & Max   & timeout\\ 
Relaying Party & Idle & Throttling          & Throttling  & Depth & $\times$ depth \\
\hline 
Routinator     & 300s & 300s                & -           & 32    & 2.6h \\  \hline 
Fort           & 24s  & 17 mins*            & -           & 31    & 9.1h \\  \hline 
OctoRPKI       & 60s  & 60s                 & 20 mins     & 30    & 30min/10h \\  \hline 
Ripe-Validator & 60s  & 0s                 & -           & $\infty$ & ($\infty$) \\ \hline
\end{tabular} \\
{
\footnotesize
*No timeout. Depends on file size and bandwidth limit set by the attacker. %
}
\caption{Stalling Results}
\vspace{-10pt}
\label{tab:stallingtest}
\end{newenv}
\end{table}

\end{newenv}

\subsection{Success Probability Analysis}\label{sc:success:analysis}

The overall success probability of our adversary depends on 2 factors: successful blocking of an individual relying party's (HTTP or DNS) request and the number of requests during the time period required for a successful attack.

{\bf The time period of an attack.} Our attack uses the fact that a relying party will switch a network's RPKI status to 'unknown' based on any break in the chain leading to the ROA for this network. Since relying parties cache the contents of RPKI repositories, the time period for an attack is therefore the shortest remaining length of any certificate in the trust-chain (i.e., often 6 hours). \new{In our analysis we consider the worst case, but in practice the manifests are not "signed on the fly" when downloaded from a publication point, hence have a shorter validity period.} For a successful attack an adversary should be able to deny connectivity to the publication point of a repository typically for at most one day, see Section~\ref{sc:attack-overview}.

{\bf Query volume.} Using the values from our relying party and stub resolver analysis  (Section~\ref{sc:analysis}), we compute the number of connection attempts during the attack period $t_{attack}$ of 1 day by multiplying with the expected number of repository updates during that period (i.e., the inverse between 2 repository synchronisation $t_{sleep}$) and the number of retries $n_{retries}$: $$n_{attempts} = t_{attack} \cdot \frac{3600s}{t_{sleep}} \cdot n_{retries}$$
We calculate the values for $n_{attempts}$ using 4 different example scenarios in Table~\ref{tab:nb_attempts}:

\textbf{(1)} An old manifest file which has only 6 hours of validity left, combined with an Unbound resolver, in case the attacker is able to deny all requests, which leads Unbound to "block" the nameserver, effectively limiting the rate of DNS requests to the nameserver to one per 15 minutes.
\textbf{(2)} A fresh manifest file which has 1 day of validity left, combined with a victim relying party using \textit{Routinator} on Linux and a Bind9 resolver. 
\textbf{(3)} A fresh manifest file with 2 days of validity left, combined with a victim relying party using \textit{RIPE NCC RPKI Validator} and an Unbound resolver where the attacker is unable to "block" the server, causing up to 16 retries per client DNS query.
\textbf{(S)} The attacker uses the \new{Stalloris} technique from Section~\ref{sc:optimization-delegation} to reduce the number of connection attempts by the RP \new{under the same conditions as scenario \textbf{(2)}.}

The first two scenarios represent the common values and configurations we measured in the RPKI setups in the Internet. The third scenario represents the theoretical worst case scenario that we defined using worst case values we collected. The last scenario uses the same values as scenario (2), but assumes the attacker uses the Stalloris optimisation. Our analysis in this section uses the number of retries in Bind9 and Unbound DNS resolvers software, as well as in Linux (the OS in RPKI deployments of relying parties and publication points) which we explored and summarised for interested reader in Appendix, Section \ref{sc:resolver-os-retries}.
\begin{table}[t!]
    \renewcommand{\arraystretch}{0.8}
{\footnotesize
    \centering
    \setlength{\tabcolsep}{4pt}
    \begin{tabular}{c|c|c|c|c|c}
    \textbf{(Scenario)} & $n_{attempts}$ & $t_{attack}$ & $t_{sleep}$ & $n_{retries}$  & o\\

             \hline
             &                & old          & unbound     & unbound        & \\
    {\bf(1)} & 24             & manifest     & (blocked)     & (blocked)        & 35 \\ 
             &                & $6$ $hours$  & $900$ $s$   & $1$            & \\
             \hline
             &                & fresh        & routinator  & bind9 /         & \\
    {\bf(2)} & 864            & manifest     & (normal)            & linux tcp      & 1247 \\
             &                & $1$ $day$    & $600$ $s$   & $6$            & \\
             \hline
             &                & long-valid   & RIPE NCC    & unbound        & \\
    {\bf(3)} & 23040          & manifest     & validator   & normal         & 33240 \\
             &                & $2$ $days$   & $120$ $s$   & $16$           & \\
             
             \hline
             &                & fresh        & routinator  & bind9 /        & \\
   {\bf(S)}& 55             & manifest     & (stalled)     & linux tcp      & 80 \\
             &                & $1$ $day$    & $2.6$ $hours$& $6$           & \\

    \end{tabular}
    \caption{Number of connection attempts in different scenarios. }
    \label{tab:nb_attempts}}
    \vspace{-10pt}
\end{table}

{\bf Success probability.} Based on the number of connection attempts and the probability to deny a connection attempt, we calculate that overall success to deny all connections during an attack period $t_{attack}$. The probability to deny a connection attempt, is:
$$ p_{connectonce} = \frac{r_{limit}}{1 + r_{attacker}} $$
where $r_{attacker}$ is the query rate of the attacker. %

We calculate the total attacker success rate by deriving the probability of denying not one but $n_{attempts}$:

$$ p_{success} = (1 - p_{connectonce})^{n_{attempts}} = (1 - \frac{r_{limit}}{1 + r_{attacker}})^{n_{attempts}} $$

{\bf Packet volume for a successful attack.} For a given success rate, we can obtain the relation between the rate limit and the required rate of spoofed requests sent by the attacker, which we call the overwhelming factor $o$: 

$$ o = \frac{1+r_{attacker}}{r_{limit}} = \frac{1}{ 1- \sqrt[n_{attempts}]{p_{success}} } $$

We calculate this factor for a target success rate of $p_{success} = 50\%$ in our 3 scenarios, which is listed in Table~\ref{tab:nb_attempts}. In the best case scenario, the attacker only needs to send roughly 35 times as many packets as the rate-limit to the publication point or nameserver, while in the worst-case scenario, this factor increases to 33240 times as many packets as specified by the rate-limit.

When combining this factor with the measured rate-limits from Sections~\ref{sec:measure-ns-ratelimiting} and \ref{sec:measure-pp-ratelimiting}, we can compute the required rate of spoofed packets. For each attack scenario, we calculate the rate $r_{attacker}$ for 3 example values for the rate-limit $r_{limit}$ listed in Table~\ref{tab:packet_volumes}. Finally we combine the required rate $r_{attacker}$ with the expected variance of the relying party update schedule up to which an attacker can predict the next query (e.g., 30 seconds in \textit{Routinator}, Figure~\ref{fig:routinator_timing}).

Our analysis shows, that depending on the rate-limit and attack scenario, the rate of spoofed packets for a successful attack ranges between 105/s and 42,813,120/s for an un-optimised attack (i.e., based only on iterative low-rate packet blocking during multiple refresh intervals), with a reasonable average-case of <74,820/s. To compare this attack volume to other recent attacks: the recently proposed DNS cache poisoning attacks \cite{man2020dns,man2021dns} require above 100K spoofed packets from the attacker in the average case for identifying the correct port and DNS TXID.

Our downgrade attack can be significantly optimised by combining the low-rate packet blocking with Stalloris (Section~\ref{sc:optimization-delegation}). For a successful Stalloris downgrade attack \new{under the same conditions as scenario \textbf{(2)}} the required rate of spoofed packets is \new{reduced by the factor 16 from 74,820pkt/s to 4,800pkt/s in the average case}. This means that even in unfavourable conditions, such as extremely high rate-limit and a resolver performing maximal number of retries, using the optimisation to stall the relying party makes the attack \new{still feasible.}  In fact, the devastating impact of the attack is disproportional to its efficiency: the adversay can with just a single such attack downgrade RPKI for all clients of a large provider (see example of DE-CIX IXP, Section \ref{sc:decix}) to a large number of Internet prefixes covered by the ROAs in the publication point that the adversary blocks.

\begin{table}[t!]
  \renewcommand{\arraystretch}{0.8}
    \footnotesize
    \centering
    \begin{tabular}{c|r|r|r}
    \makecell{\textbf{(Scenario)} \\ $o$} & $r_{limit}$ & $r_{attacker}$ & \makecell[r]{total packets\\per update} \\
  \hline

    \multirow{3}{*}{\makecell{\textbf{(1)} \\ 35}}
          & 3    &        105 &       3,150 \\
          & 60   &      2,100 &      63,000 \\
          & 1288 &     45,080 &   1,352,400 \\
    \hline
    \multirow{3}{*}{\makecell{\textbf{(2)} \\ 1247}}
          & 3    &      3,741 &     112,230 \\
          & 60   &     74,820 &   2,244,600 \\
          & 1288 &  1,606,136 &  48,184,080 \\
    \hline
    \multirow{3}{*}{\makecell{\textbf{(3)} \\ 33240}}
          & 3    &     99,720 &   2,991,600 \\
          & 60   &  1,994,400 &  59,832,000 \\
          & 1288 & 42,813,120 & 1,284,393,600 \\
          
    \hline
    \multirow{3}{*}{\makecell{\textbf{(S)} \\ 80}}
          & 3    & 240        & 7,200       \\
          & 60   & 4,800      & 103,040      \\
          & 1288 & 103,040     & 3,091,200   \\

    \end{tabular}
    \caption{Packet volume for a successful attack \new{based on analysis in Section~\ref{sc:success:analysis}}. }
    \label{tab:packet_volumes}
    \vspace{-10pt}
\end{table}

\subsection{Vulnerable Networks}

We calculate the number of networks vulnerable to removal of RPKI protection in terms of network size. We consider all network blocks vulnerable which are hosted on repositories with rate-limiting (DNS or TCP) enabled, see Tables~\ref{tab:pp_ns_rate_limits} and \ref{tab:tcp_syn_rate_limits}. We extract a mapping of publication point domains to the ROAs hosted on these publication points from a running Routinator instance to see which networks are affected by removal of RPKI protection in the case this publication point is targeted by an attacker.
We also calculate the number of networks which have RPKI protection by summing up the network sizes of all ROAs in all repositories. Our calculation does not consider other methods of Denial-of-service, such as rate-limiting in resolvers used by relying parties (See Section~\ref{sec:measure-resolver-ratelimiting}), as these methods depend on the relying party only and cannot be reliably measured in the internet without causing collateral damage.

\new{Our results are shown in Table~\ref{tab:vuln_networks}.} In conclusion we find that 34.2\% (IPv6: 40.0\%) of the assigned address space has published ROAs, out of which 59.6\% (IPv6: 37.9\%) are potentially vulnerable to removal of RPKI protection due to our attack. Furthermore, in 3.1\% (IPv6: 3.2\%) of the address space, these attacks are highly practical due to to low rate limits at or below 60 requests/sec. These cases are made up almost entirely from the customers of \new{a single provider,} which is the only major operator of a publication point which such a low rate limit\footnote{The other 2 repositories both hosted ROAs only for a small number of relatively small network blocks.}.

\begin{table}[t!]
  \renewcommand{\arraystretch}{0.8}
    \centering
    \footnotesize
    \begin{tabular}{r|c|r|r|r}
                    &    & total    & \% of assigned & \% of ROA- \\
                    &    & addresses& address space & protected  \\
    \hline
    has ROA         & v4 & ~64 * /8   & 34.2 \%  & 100.0 \%     \\
                    & v6 & ~322 * /24 & 40.0 \%  & 100.0 \%     \\
    \hline
    vulnerable      & v4 & ~39 * /8   & 20.4 \%   & 59.6 \%    \\
    (all)           & v6 & ~122 * /24 & 15.2 \%   & 37.9 \%   \\
    \hline
    vulnerable      & v4 & ~2 * /8    & 1.1 \%   & 3.1 \%     \\
    (low ratelimit) & v6 & ~10 * /24  & 1.3 \%   & 3.2 \%    %
    \end{tabular}
    \caption{Networks vulnerable to removal of RPKI protection.}
    \label{tab:vuln_networks}
    \vspace{-10pt}
\end{table}

\section{Find the Victim}\label{sc:find}
In order to attack a target AS the adversary needs to identify the relying parties that the victim AS uses for validating the ROAs. We develop a methodology for associating between an AS and a relying party that it uses in RPKI, with the conflicting ROAs that ROV measurements use \cite{gilad2017we,DBLP:conf/dsn/HlavacekHSW18,reuter2018towards,reutermeasuring}. Our approach works by selectively presenting different sets of conflicting ROAs to the relying parties while monitoring the reachability of the corresponding network blocks. We identify dependency by observing the routing decisions that the routers make based on validated ROAs computed by the relying party. %

We assume that the target network is identified by an ASN $AS^T$ and that it enforces ROV. Furthermore, the $AS^T$ announces $N^T$ IP prefixes $P^T_i$, $i \in \{1,2,\ldots,N^T\}$ that we identify using the BGP table dump\footnote{\url{https://www.ripe.net/analyse/internet-measurements/routing-information-service-ris/ris-raw-data}}.

{\bf Setup.} Carrying out the procedure for associating a relying party with IP address $a^{RP}$ to the target network $AS^T$ requires the following: A controlled network $AS^A$ (simulated attacker), a BGP router in $AS^A$ that can announce IP prefixes $P^A_1$ and $P^A_2$ to the Internet, the RPKI CA and the publication point that we control fully and a valid RPKI delegation for these IP prefixes towards our RPKI CA and PP.

{\bf Procedure.} We execute the following steps to test if network $AS^T$ depends on the relying party with IP address $a^{RP}$:

(1) Select a target IP addresses $a^T$ from one of the IP prefixes $\{P^T_1, P^T_2, \ldots, P^T_N\}$ announced by the $AS^T$. The selection of the $a^T$ requires that this IP address respond to ICMP echo requests or there is at least one publicly available TCP or UDP service on the selected IP address and it can be easily probed for reachability. We assume that at least one such IP address exists in the target network and furthermore we assume that the routing policy is uniform for all prefixes  $\{P^T_1, P^T_2, \ldots, P^T_{N^T}\}$.

(2) Announce IP prefixes $P^A_1$ and $P^A_2$ from $AS^A$.

(3) Create two sets of ROAs for the prefixes $P^A_1$ and $P^A_2$: The first set is $\Sigma = \{\rho_1, \rho_2\}$ where $\rho_1 = (P^A_1, AS^A, max\_len=pfxlen(P^A_1))$ and $\rho_2 = (P^A_2, AS^A, max\_len=pfxlen(P^A_2))$, $pfxlen(P)$ denotes the true IP prefix length of the prefix $P$. Both ROAs $\rho_1$ and $\rho_2$ are in agreement with the BGP announcements for $P^A_1$ and $P^A_2$ from the previous step. Furthermore we define the second ROA pair $\overline{\Sigma} = \{\overline{\rho_1}, \rho_2\}$ where $\overline{\rho_1} = (P^A_1, X, max\_len=pfxlen(P^A_1))$ and $X \neq AS^A$. The ROA $\rho_1$ contradicts the $P^A_1$ BGP announcement, hence the validation outcome with ROA pair $\overline{\Sigma}$ is inverse from $\Sigma$ for prefix $P^A_1$, but the second prefix $P^A_2$ is not blocked by ROV in both cases for positive reachability verification.

(4) Set up a RPKI CA along with a modified publication point that can sign both sets $\Sigma$ and $\overline{\Sigma}$ and selectively respond to RPs with the specific ROA set based on {\it RRDP} and {\it rsync} query source IP address. We set up a delegation for prefixes ${P^A}_1$ and ${P^A}_2$ to this CA and PP.

(5) We configure our PP from the previous step to respond with ROA set $\Sigma$ to all RPs except one with IP address $a^{RP}$. Only for this RP our PP responds with ROA set $\overline{\Sigma}$. Hence the RP with IP address $a^{RP}$ is the only RP that receives the inverse ROA for $P^A_1$.

(6) We define (bi-directional) reachability test using ICMP echo or the service selected for the specific destination IP address in step (1):

\begin{newenv}
$R(s, d) =\left\{
        \begin{array}{ll}
            1  & \quad \textrm{if the IP address } d \textrm{ responds to a request}\\
               & \quad \textrm{from IP address } s\\
            0  & \quad \textrm{otherwise}
        \end{array}
    \right.$

The selected IP address $a^T$ in the target network is probed from two distinct source IP addresses $a^A_1 \in P^A_1$ and $a^A_2 \in P^A_2$. The possible outcomes are:

    $\bullet$ $R(a^A_1, a^T) = 0 \land R(a^A_2, a^T) = 1$ and all intermediary networks on path between $AS^A$ and $AS^T$ do not enforce ROV or their RPs have already been identified and differs from $a^{RP}$: It indicates that $AS^T$ enforced ROA set $\overline{\Sigma}$. This set is distributed only to RP with IP address $a^{RP}$, hence the network $AS^T$ uses VRPs from this RP to implement ROV.
    
    $\bullet$ $R(a^A_1, a^T) = 1 \land R(a^A_2, a^T) = 1$: ROA set $\overline{\Sigma}$ is not enforced in $AS^T$, hence the network $AS^T$ does not use VRPs from RP with IP address $a^{RP}$ to implement ROV.
    
    $\bullet$ Any other result is considered invalid and therefore the relationship between the $AS^T$ and RP with IP address $a^{RP}$ can not be determined.

Since we fully control the publication point, we collect all the IP addresses of active relying parties and repeat the steps (5) and (6) for each active RP IP address as the $a^{RP}$. By this method we can find a RP for a specific network. %

{\bf Time complexity.}
Our measurements in lab using a modern Routinator v0.9.0 RP with a router as the RTR client (Cisco ASR 9000 running Cisco IOS XR Release 7.4) show that the changes propagate to the routers within few seconds after the refresh. However, we expect that longer waiting times between the RP setup and reachability tests are needed to counter the batch-processing nature of the ROA validation and validated ROA payloads update workflow over RTR, limiting the real speed to up to two rounds per hour.

\end{newenv}

\section{Use Case: Internet Exchange Point (IXP)}\label{sc:decix}

The IXPs provide a high-capacity L2 interconnect between the members - ISPs and networks of various sizes. Typically networks in IXPs use route servers to avoid operational, administrative and management challenges associated with setting up and maintaining the BGP sessions among themselves. In our example we use Frankfurt DE-CIX IXP, one of the largest IXPs in the world, as a test case. In DE-CIX this constitutes about 85\% of the networks\footnote{\url{https://lg.de-cix.net/routeservers/rs1_fra_ipv4}}, which were 1301 configured peers on Oct 5, 2021. 

{\bf Adversary model.} The attacker needs to be a member of the IXP, which means that it has direct BGP connection to the route servers and therefore it can effectively execute the algorithm for associating the target network with the relying party in an automated manner. In addition, the network of the adversary needs to run the delegated RPKI infrastructure - every network has the freedom to decide whether it operates a delegated RPKI infrastructure or a hosted RPKI at the RIR (Regional Internet Registry). Moreover, the attacker needs to have access to the internal and public network management and debugging tools provided by the IXP - Looking Glass service, list of members with their peer IP addresses and port state and statistics. All of these are provided to the members at the IXP.

{\bf Reliance on the route server.} The route servers in modern IXPs implement strict filtering rules\footnote{\url{https://www.de-cix.net/en/locations/frankfurt/route-server-guide}}, that include ROV. The members of the IXP prefer the routes from the IXP (either from route servers or from direct BGP sessions) due to the higher spare capacity, lower cost and direct path, in comparison with the upstream connectivity. This preference for IXP routes is implemented by {\scriptsize{LOCAL\_PREF}} attribute. Preferring the IXP routes introduces a higher risk for BGP hijacking, because {\scriptsize{LOCAL\_PREF}} has one of the higher priorities in the BGP best path selection algorithm [Section 9.1 in RFC4271]. Namely, once RPKI validation is removed the members at the attacked IXP will be redirected to the attacker's prefix. Our measurements of the peers at the DE-CIX IXP show that they do not perform checks themselves but only rely on the ROV performed by the route server at the IXP.

The adversary can easily obtain access to the core Internet infrastructure, allowing it to inject BGP routes that will be preferred the other paths (even those with higher {\scriptsize{LOCAL\_PREF}}). We explain this in Appendix, Section \ref{asc:capabilities}.

The aim of our attack is to inject and achieve a dissemination of an unauthorised route in BGP by the route server. The route we want to inject would not pass ROV validation since it violates the ROA of the victim AS (it maps the adversary's ASN to the prefix of the victim). Therefore, we first force the relying party at DE-CIX to not perform RPKI validation for the public RPKI repository that the victim AS uses.
In this attack we use our own AS as a victim with a public repository that we control.

{\bf Injecting hijacking BGP announcements into IXP.} The steps of the attack are as follows:
 (1) we locate the relying parties used by the route servers in the IXP. (2) Identify the public RPKI repositories that published the ROAs relevant for the IP prefix that we want to hijack. (3) Launch the downgrade attack to degrade the RRDP or rsync service between the PP and RP(s). (4) Wait for the manifests to run out of the validity period so the RP(s) drop all the ``suspicious'' ROAs according to [section 6.4 in RFC6486]. Finally (5) Announce the unauthorised prefix to hijack the traffic of the victim AS.

\new{
{\bf Ethical considerations.} Since Stalloris attack can stall an IXP our evaluations were coordinated against our own relying party. %
In our evaluations the publication point was targeting only the relying party of our test AS, which was connected to that IXP. The publication point provided a specially crafted certificate with multiple downstream delegations to the relying party of our AS and normal certificates without multiple delegations to all other relying parties. Hence only our "victim" relying party was affected, but not the others. %
}
\section{Countermeasures and Mitigations}\label{sc:mitigations}

We provide recommendations for resolving the core issues as well as the technical issues that allow our specific RPKI downgrade attacks. 

{\bf Validation limit on delegation chains.} The Stalloris attack can be mitigated by restricting the maximum depth and length of the delegation by enforcing a global limit on the delegation chain. Our evaluations show that 32 would be sufficient for all legitimate delegations. In addition, Asynchronous I/O (e.g., with {\tt asyncio} library) can be used to run the refresh in parallel. \new{Limiting the delegations would prevent our Stalloris optimisation to the downgrade attack, but since the low-rate downgrade attack is not implementation dependent, this would not prevent the low-rate downgrade attack.} We experimentally showed that in many cases even the non-optimised version of the low rate attack is practical.

{\bf ``Unknown'' RPKI validation status.} The core problem which exposes to downgrade attacks is that the adversaries can cause RPKI validation to result in status ``unknown''. A way to fix this would be to return status ``invalid'' in situations where ROAs cannot be located. The downside of this solution is that benign network failures or misconfigurations would prevent connectivity to the affected network prefixes.
Furthermore, because the repositories host many different ROAs, relying parties could only block the complete address space delegated to the affected publication point. In case of a RIR repository, this would affect the complete Internet since all the RIRs can generate ROAs for all network blocks.
In addition, adversaries could also exploit the strict validation to cause DoS attacks by preventing access to the repositories thereby leading to invalidation of legitimate network prefixes. It is important to evaluate and analyse the tradeoff between the permissive and the strict approaches and to identify what is more suitable for Internet routing. In addition, every network should be able to decide and configure this for their network themselves.

{\bf Distribution and redundancy.} A significant weak link is the single URL that is currently used to find the publication point. Our proposal is to create multiple NS (nameserver) records\footnote{Currently the publication point domains use approx. 3 nameservers, which is too few.} and to distribute the URLs of the publication points. This would solve not only downgrade attacks but also the benign failures. If one instance fails, you can try the others. Multiple locations for retrieving the RPKI information would introduce redundancy to avoid failures. In addition, it is important to host the publication points on robust platforms that guarantee high degree of connectivity.

{\bf Rate limiting.} The property which we used in this work for developing the packet loss technique exploits rate limiting in DNS or in publication points. Rate limiting threshold can be increased or rate limiting can be applied over all source IP addresses. This would make the attack via rate limiting more difficult to launch. We caution however, that other techniques can be used for launching our downgrade attacks, such as packet loss with spoofed fragmented IP packets \cite{KenMog87,gilad2017we} or packet loss with an adversary that controls an intermediate router and introduces selective losses into the traffic that traverses it. Blocking different methods for causing packet loss does not resolve the main problem which is inherent in the behaviour of relying party implementations. 

{\bf Randomisation of refresh interval.} We recommend that the refresh interval of the relying party implementations is randomised. This would make our attacks difficult to launch for MitM adversaries and almost impractical for off-path adversaries, since the point in time when the attack needs to be launched would be impossible to predict. Hence, to launch our downgrade attack the adversary would need to constantly prevent access to the repository, e.g., by continually flooding the communication. This however, exposes the attack.

{\bf Eliminating manifests.} Eliminating manifests and using the actual validity of the ROAs could make it more difficult to launch the attack since the adversary would need to wait longer for RPKI objects to expire. The downside of this is that eliminating the manifests would increase the revocation lists and any changes in ROAs would create a lot of updates. This would be especially significant for large providers, such as RIRs, which host thousands of networks, and would result in huge revocation lists. Manifests were introduced to avoid long revocation lists - the relying party has to update only the manifests but not the complete set of ROA objects.

\section{Related Work}\label{sc:works}

{\bf Deployment of RPKI.} Gradually the deployment of RPKI is increasing as more and more ASes create ROAs and start enforcing ROV to filter bogus BGP announcements. Measurements of ROA objects and of ROV filtering show a stable increase over the last decade, \cite{iamartino2015measuring,gilad2017we,rovdsn2018,clark2020filter}, with occasional failures in retrieving the ROAs \cite{kristoff2020measuring}. A recent proposal \cite{hlavacek2020disco} automated the manual certification of IP prefixes, essentially resolving the last obstacle towards large scale adoption of RPKI. %

Initial deployments of RPKI were characterised by a large fraction of erroneous ROAs \cite{gilad2017we} that not only did not guarantee security but worse, ROV filtering of such ROAs cause networks to lose legitimate traffic. Gradually the misconfigurations are being resolved, and since 2018 an increasing number of networks are enforcing ROV \cite{clark2020filter}.

\new{The increased adoption of RPKI motivates closer inspection of the security of RPKI deployments. In this work to create our dataset of ASes we collect the networks that deployed RPKI. But in our work this is not the end, but only the first step. We then evaluate vulnerabilities of the RPKI deployments to downgrade attacks.}

{\bf Vulnerabilities in RPKI.} Limitations of RPKI as well as possible misconfigurations were considered in previous work. Research of NTP (Network Time Protocol) ecosystem \cite{malhotra2016attacking} suggested, without validating in practice, that shifting time may affect different systems, including RPKI, by causing the relying party to accept stale manifest as valid. \new{In this case RPKI validation is performed over a stale manifest.} Another concern with RPKI is the deliberate or accidental exposure to IP prefix takedowns which causes prefix of the affected AS to become unreachable \cite{cooper2013risk}. \new{A malicious registrar can issue a certificate to invalidate a victim prefix, hence causing other ASes to filter its traffic.} Finally, recently \cite{morillo2021rov++} showed that partial ROV deployment has only limited security benefits and proposed an extension to ROV providing benefit against sub-prefix hijacks also for early adopters. \new{In our work we find vulnerabilities in RPKI implementations and demonstrate attacks against RPKI deployments in the Internet. Our attacks apply even if RPKI is widely supported by all the ASes. Our adversary is weak, it can only send packets from a spoofed IP address but does not control a registrar.}
 
\begin{newenv}
Our attacks combine rate limiting at the servers with low rate bursts of packets to cause loss of requests sent by the relying parties. Low rate bursts were initially demonstrated to cause practical reset of TCP connections \cite{conf/sigcomm/KuzmanovicK03}. We then show how to optimise our low rate attacks with Stalloris attack. In a Stalloris attack we set up a malicious publication point or a CA, that exploits certain properties in relying parties, causing them to stall.

The name of our attack draws from Slowloris \cite{cambiaso2013slow}, which is a DoS attack against HTTP servers in which a malicious client connects to the victim server by sending partial HTTP requests at a exceptionally slow rate (only sending a few bytes) in an attempt to exhaust the amount of HTTP connections a server can hold open at once. In a Stalloris attack, the clients (relying parties) are the victims and the publication points are malicious. The main aspect of our attack is to slow down the relying parties by creating complex delegations which cause the relying parties to stall. This attack can also be enhanced by sending slow HTTP or rsync responses. In addition, in contrast to Slowloris, our goal is not to deny service to clients, but to slow down the relying parties so that they do not finish their task.
\end{newenv}

\section{Conclusions}\label{sc:conc}

The key prerequisite for delivering Internet services is connectivity and RPKI is no exception to this. When RPKI cannot retrieve an ROA to validate a route in BGP, a relying party can lose connectivity to the IP addresses in the missing ROA. The design choice that the inventors of RPKI made was to give up the RPKI security for connectivity, introducing the ``unknown'' status for RPKI validation, which allows to accept routes without requiring a valid ROA. Our work however shows that it is important to find the correct balance between connectivity and security: RPKI is vulnerable to downgrade attacks.

We develop a Stalloris RPKI downgrade attack, that inflicts packet loss in a specific time interval synchronised with the refresh interval of the relying parties in combination with an attack that stalls the relying party. This causes the relying party to give up the RPKI validation when performing routing decisions in BGP. We devise an approach for inferring the refresh interval, which makes the downgrade attacks practical not only for MitM, but also for off-path adversaries. We demonstrate RPKI downgrade attacks with off-path adversaries by exploiting the rate limiting mechanism in servers in the RPKI infrastructure. We show experimentally that 60\% of the RPKI protected networks are vulnerable to our downgrade attacks with rate-limiting packet blocking. We further find that 47 (77\%) of the vulnerable publication points are also vulnerable to sub-prefix hijack attacks. We caution that the extent of the vulnerabilities is potentially larger since other methods can be exploited for causing packet loss remotely. Full adoption of RPKI does not prevent our attacks.

We recommend countermeasures to mitigate the threat. Nevertheless our work shows that finding a balance between connectivity and security is a challenging problem.

    \new{
\section*{Acknowledgements}
We thank the reviewers for their helpful comments on our work. This work has been co-funded by the German Federal Ministry of Education and Research and the Hessen State Ministry for Higher Education, Research and Arts within their joint support of the National Research Center for Applied Cybersecurity ATHENE and by the Deutsche Forschungsgemeinschaft (DFG, German Research Foundation) SFB~1119.
}
\pagebreak
{\small{
\bibliographystyle{plain}
\bibliography{sec,NetSec}
}}

\appendix
\section{Obtaining Access to Infrastructure}\label{asc:capabilities}
IXP are open and anonymous places due to scale (European neutral IXPs have hundreds or even over a thousand members) and there is a fierce competition drive prices for entry-level ports to low or even symbolic prices, like 1 EUR for 1 Gbps port (https://www.nix.cz/en/services). There are resellers that offer VMs with the direct connectivity to several IXPs ( https://ifog.ch/en/ip/ixp-access ), thus the attacker can obtain access to the core Internet infrastructure that allows him to inject BGP routes over the generally-preferred path (with higher LOCAL\_PREF) for prices as low as 25 EUR/month (plus a price for allocation/renting/stealing ASN). We consider this environment to be a new attack surface on BGP. The route-server side checks and ROV are therefore needed and the IXP members depend on it to high extent (most of the detected ROV adopters (see https://rov.rpki.net/ ) do not implement it internally, they are protected by a ROV-enforcing route server in an IXP.

\section{Retries in Other Components:\\Resolver and OS-level}\label{sc:resolver-os-retries}

Additionally to any application-level retries, we analyse the number of network packets emitted by a relying party or the resolver it is using.

{\bf Retries in DNS resolvers.} We analyse the retransmission behaviour in the two most commonly used recursive DNS resolvers: Bind9 and Unbound. In bind Bind9, the number of retries sent upon not receiving an answer to a DNS query is controlled by the overall client query timeout and the individual timeouts after which a packet loss is assumed \cite{bind-docs}. By default, the overall timeout is 10 seconds and the time between queries starts at 800ms and is doubled after the forth unanswered query. This results in 6 queries during the 10 second period.

In unbound, the behaviour is more sophisticated \cite{timeout-unbound}: while Unbound starts by sending up to 16 queries to a nameserver in our tests, upon not receiving answers from a nameserver it marks the nameserver as being unresponsive. This process has two levels and on its highest level, where a nameserver is marked as ``blocked'', Unbound will only allow 1 query every 15 minutes to that individual nameserver and will answer all other client queries with SERVFAIL immediately until an answer from this nameserver is received again or the information that the nameserver was blocked is cleared from the cache after 15 minutes.

{\bf Retries in TCP.} We also analyse the retransmission behaviour for TCP connection in the case the attacker is performing the rate-limiting attack directly against the publication point. In linux the amount of TCP SYN packets is controlled by the \path{tcp_synack_retries} parameter \cite{linux-tcpretry}, which is set to 5 by default, which results in 6 packets in total. In windows, this number is controlled by a \path{TcpMaxConnectRetransmissions} registry setting \cite{windows-tcpretry}, which is set to 2, meaning a number of 3 packets. Since most relying party software is build for linux systems, we only take this number into account in our calculations.

\begin{newenv}
\section{RPKI Overview}\label{sc:background}

{\bf BGP prefix hijacks.} In a BGP prefix hijack an adversary creates a bogus BGP announcement that maps the prefix of the victim AS to its own ASN (AS number). As a result the adversary can intercept traffic from all the ASes that have less hops (shorter AS-PATH) to the attacker than to the victim AS. The same-prefix hijack affects the traffic of the ASes that prefer the attacker's announcement. The effectiveness of the same-prefix hijack attacks depends on the local preferences of the ASes and the location of the attacker's AS. The adversary can also advertise a sub-prefix of the victim AS's prefix. The routers prefer more specific IP prefixes over less specific ones, hence the longest-matching prefix (e.g., /24) gets chosen over the less-specific prefix e.g., /20). Once a victim AS accepts the hijacking announcement it sends all the traffic for that sub-prefix to the adversary.

{\bf RPKI.} RPKI associates public keys with IP prefixes [RFC6480]. After certifying their IP prefixes, owners can use their private keys to sign Resource Certificates (RCs) and Route Origin Authorizations (ROAs), which authorise AS numbers to advertise these prefixes in BGP. RCs and ROAs are published on publication points (aka RPKI repositories), which the relying parties periodically query to retrieve the RPKI objects. The BGP routers then use the RTR protocol [RFC8210] to fetch the validation results from the relying parties parties and apply ROV filtering for routing decisions in BGP.

RPKI supports a delegated and a hosted model. In the delegated RPKI model, AS runs a CA as a child of RIR (or NIR or LIR), generates its own certificate, gets it signed by the parent CA. This model allows the AS to operate independent of the parent RIR. For large operators of a global network, this model is suitable so that they do not need to maintain ROAs through the different web interfaces of the RIRs. This model allows the owner to run his own CA, control the publication point and maintain the ROAs. In the hosted-RPKI model, RIRs host the CA, that is, the same entity that allocates IP resources also runs the CA to validate the ROAs.

To find the publication servers the relying parties use the DNS resolvers to lookup the hostnames of the repositories. The relying party software starts at the top of the cryptographically-signed chain that begins with the Trust Anchor Locator (TAL). TAL contains the URLs of RRDP and/or rsync servers and the fingerprint for download and validation of the top level CA certificate.

Each of the five RIRs operate one top level RPKI CA for the resources managed by that RIR, hence all the relying parties need the current TAL for the RPKI CA of each RIR - it has to be supplied by the operator or packaged with the relying party software. Starting from TAL, the relying party recursively contacts the publication points that form subtree of the root CA. An RPKI certificate can delegate the resources to a child publication server. The relying party traverses the trees, downloads the RPKI certificates, validates them along with other supplementary objects, manifests, CLRs and ROAs. ROA is the leaf in the RPKI tree that holds the cryptographically signed triplets that constitute the final output that the relying party provides to the border routers. 

\end{newenv}

\end{document}